\documentclass[preprint2]{aastex}

\usepackage[]{natbib}
\usepackage{graphicx}
\usepackage{amssymb}
\usepackage{color}
\usepackage{ulem}

\input{macro.inp}

\newcommand{\myemail}{Kenji.Hamaguchi@nasa.gov}

\shorttitle{Resolving a Protostar Binary System}
\shortauthors{Hamaguchi et al.}

\slugcomment{arXiv}

\begin{document}

\title{
Resolving a Class I Protostar Binary System with Chandra
}

\author{Kenji Hamaguchi\altaffilmark{1,2}, Minho Choi\altaffilmark{3}, Michael F. Corcoran\altaffilmark{1,4}, Chul-Sung Choi\altaffilmark{3}, Ken'ichi Tatematsu\altaffilmark{5}, Rob Petre\altaffilmark{6}}

\altaffiltext{1}{CRESST and X-ray Astrophysics Laboratory NASA/GSFC,
Greenbelt, MD 20771}

\altaffiltext{2}{Department of Physics, University of Maryland, Baltimore County, 
1000 Hilltop Circle, Baltimore, MD 21250}

\altaffiltext{3}{International Center for Astrophysics, Korea Astronomy and Space Science
Institute, Hwaam 61-1, Yuseong, Daejeon 305-348, South Korea}

\altaffiltext{4}{Universities Space Research Association, 
10211 Wincopin Circle, Suite 500, Columbia, MD 21044}

\altaffiltext{5}{National Astronomical Observatory of Japan, 2-21-1 Osawa, Mitaka,
Tokyo 181-8588, Japan}

\altaffiltext{6}{Astrophysics Science Division, NASA Goddard Space Flight Center,
Greenbelt, MD 20771}

\altaffiltext{}{{\tt Mail to:} \myemail}

\begin{abstract}
Using a sub-pixel event repositioning technique,
we spatially resolved X-ray emission from the infrared double system IRS~5
in the R Corona Australis molecular cloud with $\sim$0\FARCS8 separation.
As far as we know, this result - obtained from 8 Chandra archival observations between 2000 and 2005 - is the first X-ray study of individual sources in a Class I protostar binary system with a projected separation of less than 200~AU.
We extracted light curves and spectra of the individual sources using a
two-dimensional image fitting method.
IRS~5a at the south,
the source which was brighter in the near-infrared, showed three X-ray flares lasting $>$20~ksec,
reminiscent of X-ray flares from pre-main sequence stars, 
while the northern source (IRS~5b) was quiescent in X-rays in all the observations
except for a 2005 August 9 observation with a factor of $\sim$2 flux enhancement.
In quiescence, these sources showed almost identical X-ray spectra, with
\NH\ $\sim$4$\times$10$^{22}$~\UNITNH, \KT\ $\sim$2~keV,
and log \LX\ $\sim$30.2$-$3~\UNITLUMI.
IRS~5a showed plasma at temperatures up to \KT\ $\sim$5$-$6~keV during flares,
while the column density of IRS~5b increased by a factor of 2 
during an observation on 2005 August 9.
We discuss the evolutionary stages and variation of the X-ray activity of these sources.
\end{abstract}


\section{Introduction}
\label{sec:introduction}

Many, if not most, stars are born as a member of a multiple system through fragmentation 
of the parent molecular cloud. Because young stellar objects (YSOs) that form as a 
binary or multiple system are of the same age and chemical composition, any differences in the 
level of their stellar activity should be due to differences of stellar initial parameters such as
mass, net angular momentum and/or the presence of the companion star
(See reviews about multiplicity of young stars in \citealt{Duchene2007,Monin2007},
and magnetic activity on various stellar parameters
in ex. \citealt{Chabrier2006,Dobler2006}).
YSO binaries are therefore a useful probe to test models of the dependence of the
activity on stellar parameters for stars of the same age.

The R Corona Australis (R CrA) molecular cloud has rich association of YSOs
with various evolutionary stages \citep{Taylor1984}.
Among them, IRS~5 (a.k.a. TS 2.4) suffers strong extinction (\AV\ $\sim$45 mag)
and is classified as a Class~I protostar
\citep[][see therein for the spectral classification of YSOs]{Wilking1986,Wilking1992}.
High angular resolution near-infrared imaging separated IRS~5 
into two stellar components,
IRS~5a ($K \sim$10.9$^{m}$) and IRS~5b ($K \sim$11.5$^{m}$).  The
separation of the two stars is less than an arcsecond \citep{Chen1993,Nisini2005}. In what follows the term IRS~5 means the entire binary system,
or is used when the emitting source was unknown due to limited imaging capability.
The southern source, IRS~5a, has a stellar spectral type of K5-K7V
and is located in a low mass protostar phase on the HR diagram
\citep{Nisini2005}.
Stellar parameters of IRS~5b are poorly known due to its faintness.

IRS~5 is also known as one of the rare Class I protostars
showing strongly variable cm continuum radio emission \citep{Suters1996}
with significant circular polarization \citep{Feigelson1998}.
Such gyrosynchrotron radio emission is 
believed to be the major radio emission mechanism
of weak-lined T-Tauri Stars \citep[wTTSs, e.g.][]{Carkner1997}, 
but the emission from protostars is supposed to be absorbed
by the surrounding ionized materials heated by mass accretion.
IRS~5 does not currently show strong mass accretion activity,
but there is evidence of an outflow possibly driven by IRS5 \citep[HH~731,][]{Wang2004}.
The near-infrared spectrum of IRS~5a shows small continuum excess
above the photospheric flux,
which also suggests a low accretion rate \citep{Nisini2005}.
The infrared spectrum also shows characteristics of scattered emission.
From these results, \citet{Nisini2005} questioned
the idea that the IRS~5 system is a bona-fide protostar, 
but it may be close to the TTS phase.

IRS~5 also showed strongly absorbed thermal X-ray emission 
\citep[\NH\ $\sim$4$\times$10$^{22}$~\UNITNH, \KT\ $\sim$2~keV,][]
{Koyama1996,Neuhaeuser1997,Forbrich2006,Forbrich2007,Forbrich2007b}.
The emission was highly variable, accompanied by occasional rapid flares.
These X-ray characteristics are consistent with plasma heating by magnetic activity 
in the stellar corona, so that the
X-ray and radio emission from IRS~5 is suspected to arise from the same root cause.
Similarly,
\citet{Forbrich2007} did not find a clear correlation between the X-ray and near-infrared 
flux variation from IRS~5 and other cluster members, which 
tends to support a coronal origin for the X-ray activity
instead of accretion driven activity.

To understand the relation between the X-ray and radio activity of IRS~5 and
study the evolution of the activity,
we need to resolve the IRS~5 binary system in both wavelengths.
\citet{Choi2008} successfully resolved IRS~5 into two cm radio sources
in a VLA radio observation.
Resolving X-ray sources with a sub-arcsecond separation is challenging even
with the \CHANDRA\ observatory with its unprecedented X-ray angular resolution.
However, encouraged by a marginal resolution of IRS~5 in a \CHANDRA\ image
\citep[see Figure 5 of][]{Forbrich2007},
we revisited \CHANDRA\ archival observations of IRS~5 
and clearly resolved the individual binary components using enhanced spatial information
obtained with the Sub-pixel Event Repositioning (SER) technique
\citep{Tsunemi2001,Li2003,Li2004}.
Using this technique we were also able to extract light curves and spectra of each source 
with a two-dimensional image
fitting method
and measured the plasma parameters of individual binary stars for the first time.

The distance to the cloud is controversial.
\citet{Marraco1981} estimated it to be 129~pc from uvby$\beta$ photometry
of three stars in the cloud.
Assuming \AV~=3.1 mag,
\citet{Casey1998} derived the same distance (129$\pm$11~pc) to the eclipsing binary TY~CrA, 
which is suspected to be a cluster member.
However, \citet{Knude1998} measured $d =$ 170~pc
from the interstellar reddening of stars in the cloud having \HIPPARCOS\ distances and Tycho
B - V colors, and whose distance we use for IRS~5.

\begin{deluxetable}{lrcccccc}
\tablecolumns{8}
\tablewidth{0pc}
\tabletypesize{\scriptsize}
\tablecaption{Journal of the \CHANDRA\ Observations\label{tbl:obslogs}}
\tablehead{
&&&&\multicolumn{2}{c}{Image Shift}\\
\colhead{Abbreviation}&
\colhead{Seq. ID}&
\colhead{Observation Start}&
\colhead{Exposure}&
\colhead{$\Delta$ (R.A., Dec.)}&\colhead{StdDev}&\colhead{Off-axis}
\\
&&&\colhead{(ksec)}&\colhead{(arcsec)}&\colhead{(mas)}&\colhead{(arcsec)}
}
\startdata
CXO$_{\rm 001007}$&19&2000 Oct 7 17:02&19.7&(0.40, 0.96)&128&59.3\\
CXO$_{\rm 030626}$&3499&2003 Jun 26 12:58&37.6&(0.56, 0.28)&81&90.0\\
CXO$_{\rm 040617}$&4475&2004 Jun 17 23:17&19.9&(0.19, 0.21)&117&114.2\\
CXO$_{\rm 050808}$&5402&2005 Aug 8 2:38&15.2&(0.07, 0.06)&88&72.1\\
CXO$_{\rm 050809}$&5403&2005 Aug 9 2:39&15.0&(0.06, 0.11)&109&72.4\\
CXO$_{\rm 050810}$&5404&2005 Aug 10 1:58&15.0&(0.08, 0.06)&67&72.5\\
CXO$_{\rm 050812}$&5405&2005 Aug 12 3:13&15.0&(0.06, 0.12)&144&72.5\\
CXO$_{\rm 050813}$&5406&2005 Aug 13 1:51&14.6&(0.03, 0.01)&189&72.6\\
\enddata
\tablecomments{
Seq. ID: sequence identification number of each observation.
Image Shift: shift of the X-ray sky coordinate frame from the original X-ray data to the 2 MASS frame.
StdDev: standard offset deviation of the X-ray source positions from the 2MASS source positions in milli-arcsecond.
Off-axis: off-axis angle of IRS~5 from the nominal point.
}
\end{deluxetable}

\section{Observations}

We found eight archival pointings with \CHANDRA\ \citep{Weisskopf2002} 
that include IRS~5 in the {\FOV} (Table~\ref{tbl:obslogs}).
These observations occurred between 2000 October and 2005 August 
and used the Advanced CCD Imaging Spectrometer with 
the Imaging array (ACIS-I) at the prime focus with exposure times between 15$-$40~ksec.
During the observations,
IRS~5 was put at 1\ARCMIN$-$2\ARCMIN\ off-axis, where the \PSF\ quality is negligibly
different from the on-axis position.
We use the convention that individual \CHANDRA\ observations are designated CXO,
subscripted with the year, month and day of the observation.
We downloaded the data from the HEASARC archive\footnote{http://heasarc.gsfc.nasa.gov/W3Browse/} and reprocessed them with the CIAO\footnote{http://cxc.harvard.edu/ciao/} analysis software
version 3.4 (CALDB ver. 3.3.0.1).
We reprocessed the data, removed the pixel randomization and
utilized the SER algorithms to improve the spatial resolution.

Photon pile-up can cause source count rates to be underestimated, spectral shapes to appear harder, 
and degrades event position determination in the SER method.
Fortunately, 
observed X-ray photon counts of IRS~5 ranged between 0.02$-$0.08 \UNITCPS,
corresponding to small pile-up fractions of $\sim$2$-$8~\%\footnote{http://heasarc.gsfc.nasa.gov/Tools/w3pimms.html}.
We thus did not correct for any pile-up effect in our analysis.

\section{Analysis and Results}
\label{sec:anaresult}

\subsection{Absolute Astrometry}

This study crucially requires absolute astrometry of the X-ray images 
with a few tenths of arc-second precision.
We therefore measured sky positions of X-ray sources around IRS~5 
and cross-correlated them with sources in the Point Source Catalog of the 
2MASS All-Sky Data Release.

There is up to 7~year time span between the interval when the 2MASS observations of the southern sky 
(1998 March $-$ 2001 February) and the \CHANDRA\ observations (2000 Oct $-$ 2005 Aug).
The cluster members can move significantly within 7~years if their proper motions are suitably large.
The \HIPPARCOS\ catalogue has 4 sources with proper motion measurements \citep{Perryman1997b}.
Among them, R CrA and V709~CrA have proper motion of $\sim$80$-$100~mas yr$^{-1}$,
but they have large uncertainty in the parallax measurement ($\sim$68, 140 mas)
and almost opposite directions in their proper motions.
This would suggest that their measurements are less reliable,
possibly due to contamination of surrounding nebular emission.
Two other sources, TY CrA and HD~176386, with better parallax measurements
have small proper motions ($\lesssim$20~mas~yr$^{-1}$).
Our tentative measurement of the proper motion of cluster members 
using VLA data between 1997 and 2005 also
derived similarly small proper motions of $\lesssim$28 mas~yr$^{-1}$\citep{Choi2008}.
These results suggest that the proper motion of the cluster members 
is small, less than $\sim$150~mas in 7 years ($\lesssim$1~pixel in the Figure 1 image).

\begin{figure*}
\begin{center}
\epsscale{1.3}
\plotone{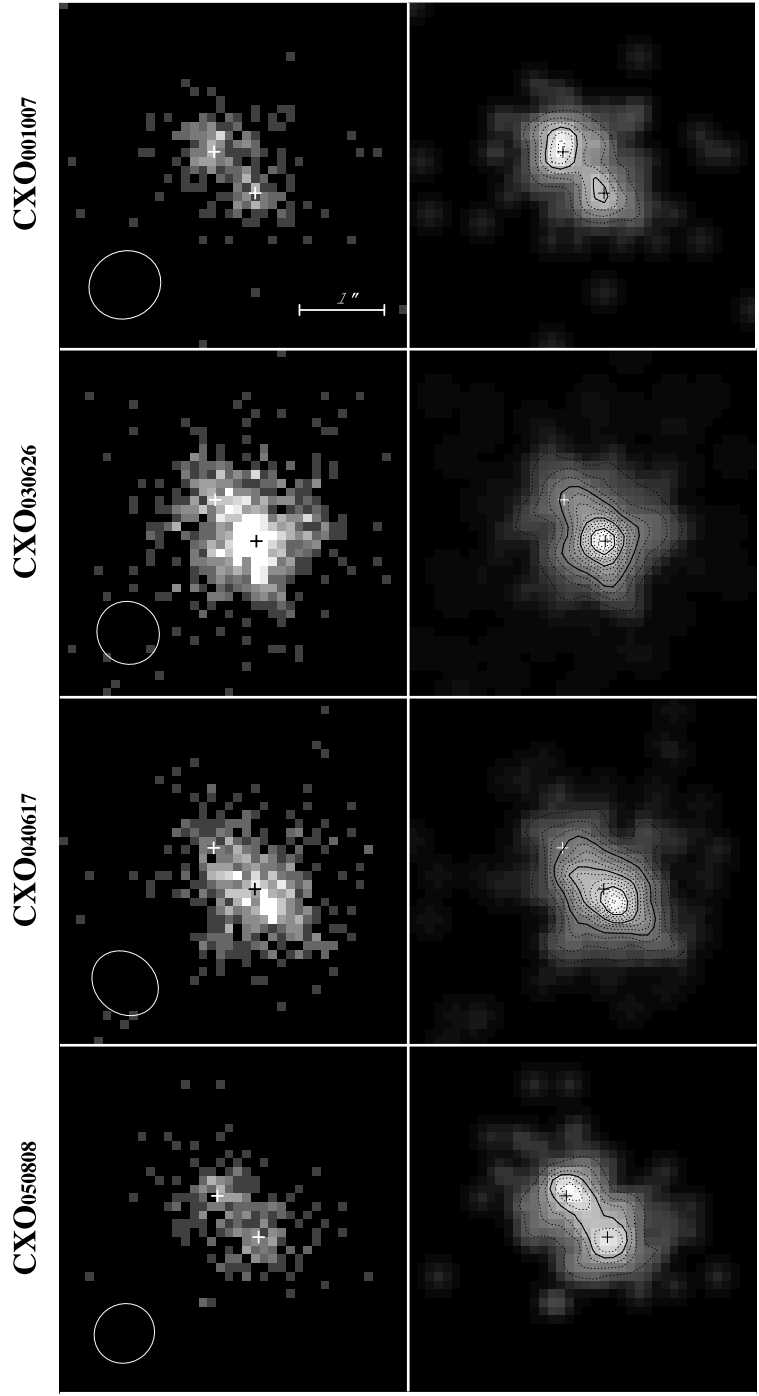}
\caption{Magnified images of IRS~5 (raw: {\it left}, 
smoothed: {\it right}), sequentially in time from above.
The gray scales are shown in a linear scale for all figures (but saturated
above 10 cnts for the left panels).
Crosses show X-ray source positions determined from our analysis.
Circles at the bottom left of the left panels show the half of the
peak intensity of the \PSF.
Raw image bins are 0\FARCS123 pixel$^{-1}$.
Gray scales and contours in the right panels 
are normalized by the effective area and exposure time in CXO$_{050808}$.
\label{fig:image}}
\end{center}
\end{figure*}

\begin{figure*}
\epsscale{1.3}
\plotone{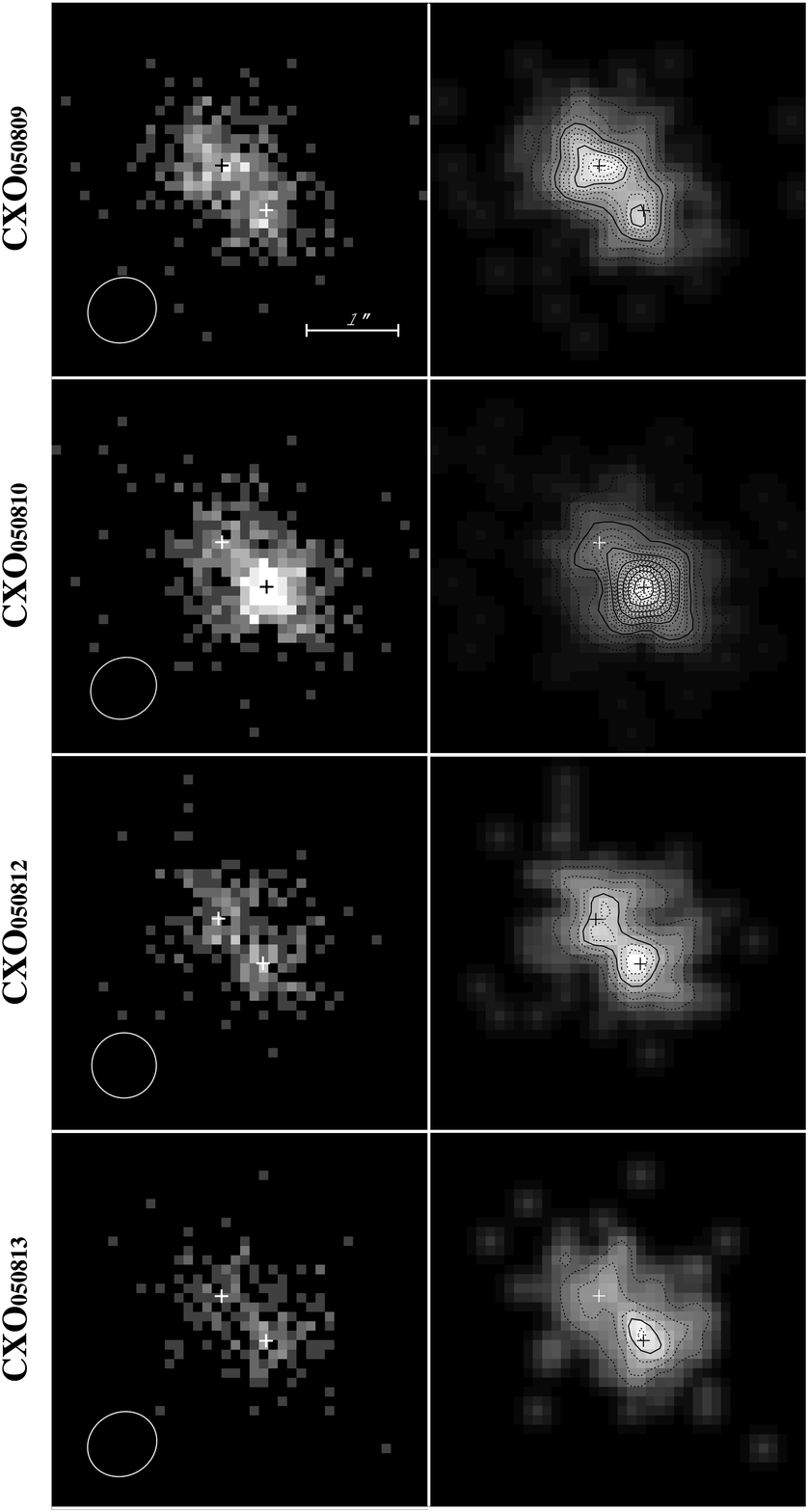}
\centerline{Fig. 1. --- Continued.}
\end{figure*}

\begin{figure}[t]
\begin{center}
\epsscale{1.0}
\plotone{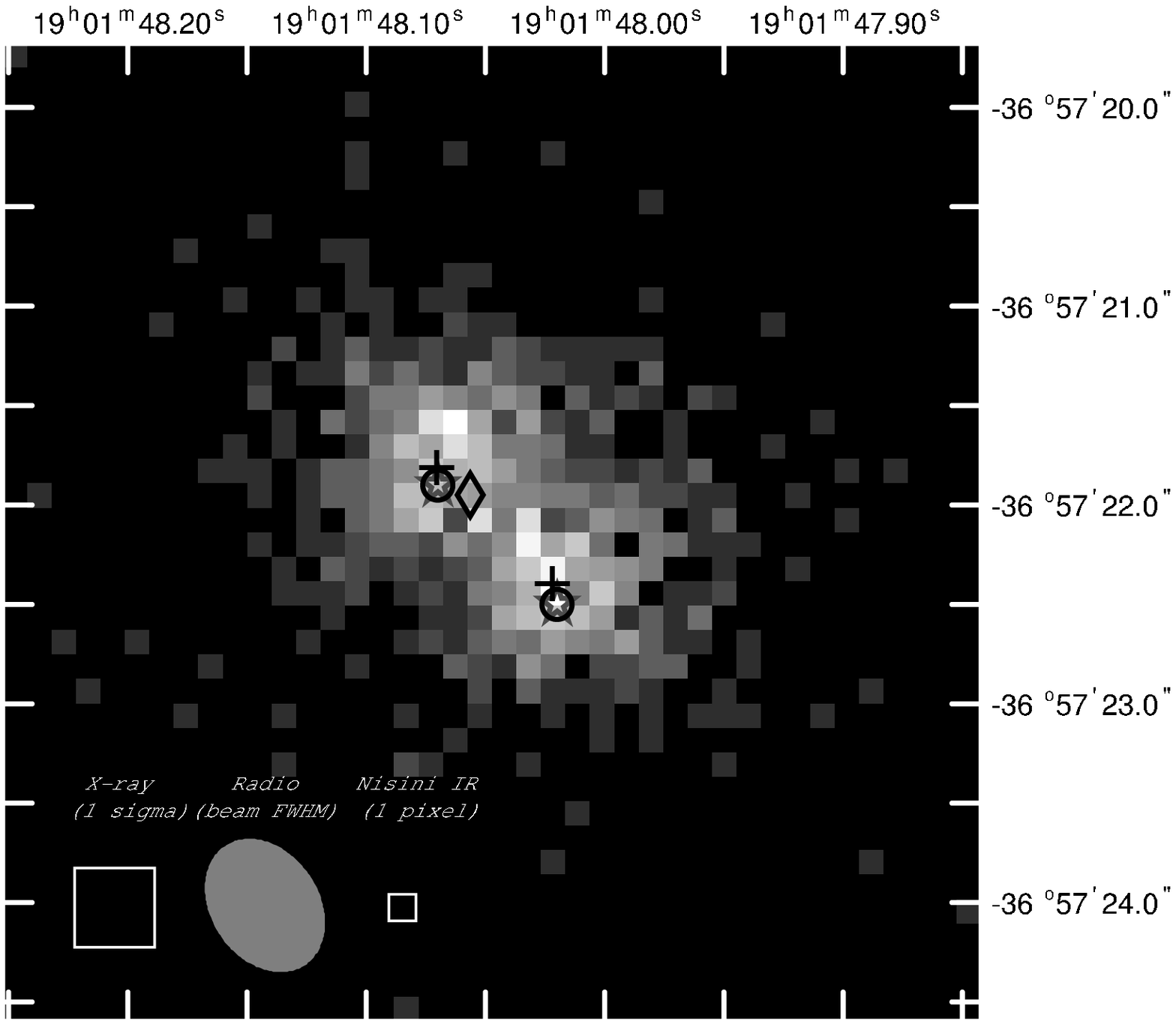}
\caption{Composite of images in CXO$_{001007}$, CXO$_{050808}$, 
CXO$_{050812}$ and CXO$_{050813}$, when both of IRS~5a and IRS~5b
were in the quiescent phase.
Markers --- cross: X-rays, circle: cm radio \citep{Choi2008},
diamond: 2MASS, star: infrared source positions in \citet{Nisini2005},
shifted to match for the cm radio source positions.
Error ranges of the X-ray source positions from the 2MASS frame (200~mas), 
radio beam size \citep{Choi2008} and IR image pixel size \citep{Nisini2005} 
are shown at the bottom left.
\label{fig:imgcomposite}
}
\end{center}
\end{figure}

To measure X-ray source positions in each observation,
we ran the {\it wavdetect} source detection tool in the CIAO package for 
a 0.35$-$8~keV image binned at 0\FARCS5 pixel$^{-1}$ and
only considered sources at above the 4$\sigma$ significance.
The 2MASS positions of IRS~1, IRS~2, R~CrA, IRS~13 and HBC~667 had
corresponding X-ray counterparts 
with similar offsets 
($\lesssim$0\FARCS1 between the sources,
see source positions in Figure~1 of \citealt{Hamaguchi2005b} and \citealt{Forbrich2006}),
while there was no counterpart to the X-ray source corresponding to IRS~6 
in the 2MASS catalogue within 1\ARCSEC.\footnote{It was in the ``reject" table.}
We therefore used the former 5 sources for the positional references.

The 2MASS positions of IRS~1, IRS~2, R~CrA, IRS~13 and HBC~667 ($K \sim$7.0--10.5$^{m}$) 
have possible position inaccuracy of up to 100~mas RMS, except for the brightest source R CrA 
($K$ =2.9$^{m}$),
which had a slightly larger position uncertainty of 160~mas (RMS)\footnote{http://www.ipac.caltech.edu/2mass/releases/allsky/doc/}.
The uncertainty of the mean position of all the sources is $\sim$50~mas.
Combined with the cluster proper motion noted above,
we can determine the absolute source position better than $\sim$200~mas for these sources.
Table~\ref{tbl:obslogs} lists the shifts of the X-ray sky coordinate frame
from the original X-ray data.
Earlier observations in 2000$-$2003 needed relatively large shifts ($\sim$0.3$-$1\ARCSEC), 
while observations in 2005 needed almost negligible shifts ($\lesssim$0.2\ARCSEC).
Table~\ref{tbl:obslogs} also lists the standard offset deviation of the X-ray source positions
from the 2MASS source positions.
The result suggests that X-ray source positions in all the \CHANDRA\ observations
were determined at $\lesssim$200~mas from the 2MASS
frame, and hence $\lesssim$400~mas in the absolute coordinate frame.

\begin{figure*}
\begin{center}
\epsscale{2.1}
\plotone{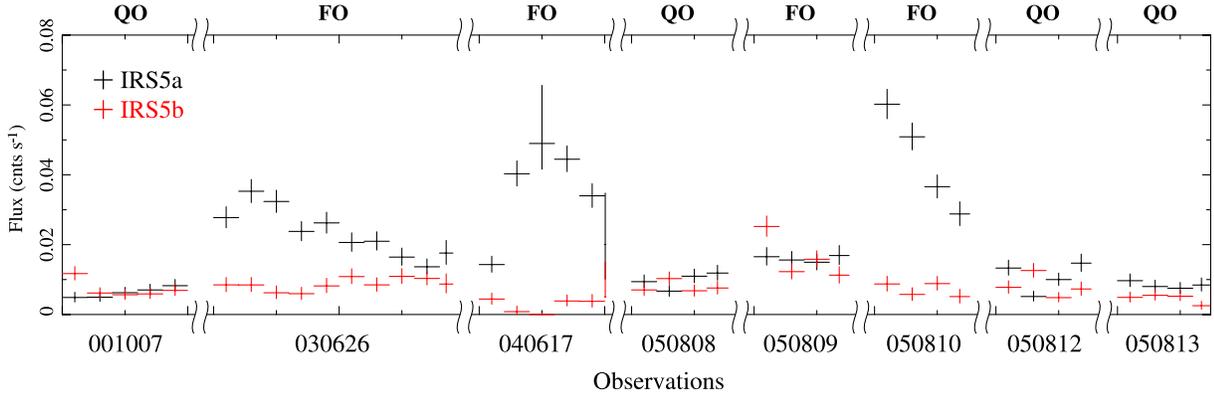}
\caption{Light curves between 1$-$8~keV ({\it top}) of IRS~5a ({\it black}) and IRS~5b ({\it red}).
Each bin has 4~ksec and tic marks on the horizontal axis are shown by 10~ksec
from the beginning of each observation.
The error bars show 1$\sigma$ confidence range.
In CXO$_{030626}$, CXO$_{040617}$, CXO$_{050810}$, and CXO$_{050813}$,
IRS~5b emitted X-rays comparable to the other observations, but {\tt wavdetect} did not detect it.
This is probably because enhanced X-ray emission from IRS~5a contaminated strongly at the 
IRS~5b position, and thus increased the apparent background level (see \S\ref{subsec:timspec}).
Labels at the top depict
observations which showed significant X-ray variations in either of IRS~5a nor IRS~5b 
(FO: Flare Observation) or in neither of them (QO: Quiescent Observation).
\label{fig:lightcurves}}
\end{center}
\end{figure*}

\subsection{Image Analysis}
\label{subsec:anaimage}

Figure~\ref{fig:image} shows X-ray images of IRS~5 between 1$-$8~keV
after the absolute position correction.
They were binned with 0\FARCS123 pixel$^{-1}$ to resolve sources having sub-arcsecond separation.
In all the images, X-ray emission was clearly elongated toward the NE-SW direction.
We checked in two ways if this structure is real.  First, we created
simulations of 
point spread functions (\PSF) at the location of IRS~5 with {\tt ChaRT} \citep{Carter2003} and 
{\tt MARX}\footnote{http://space.mit.edu/CXC/MARX/} version 4.2.1, assuming a spectrum
\NH\ =4.5$\times$10$^{22}$~\UNITNH, \KT\ =3~keV and $Z$ =0.3~solar,
(based on our preliminary fits of the IRS~5 spectra using the standard method).  Second, we made images of
IRS~1 with a similar off-axis angle to IRS~5
to determine if the SER method might introduce asymmetries in the \PSF.
We found no such extension from either of the half maximum \PSF\ circles 
(Figure~\ref{fig:image}) nor the IRS~1 images.
We conclude that the deduced elongation of the X-ray image of IRS~5 is real.

In CXO$_{030626}$, CXO$_{040617}$ and CXO$_{050810}$
the south-west part of the elongation was remarkably bright,
while in CXO$_{050809}$ the north-east part looked slightly brighter than the south-west part.
In the other observations, we recognized apparently two peaks with similar intensities.
We re-ran the {\tt wavdetect} source detection tool for these high resolution images
and detected the south-west source in all observations, and detected the north-east source in 
CXO$_{001007}$, CXO$_{050808}$, CXO$_{050809}$ and CXO$_{050812}$.
The north-east peak was not detected
when the south-west peak was very bright or images lacked sufficient exposure.
Both peaks were located at, on average, 
($\alpha_{2000}$, $\delta_{2000}$) = (19$^{h}$1$^{m}$48$.^{s}$07, $-$36\DEGREE57$'$21\FARCS8)
designated using the \CHANDRA\ naming convention as CXOU J190148.1-365722 and
($\alpha_{2000}$, $\delta_{2000}$) = (19$^{h}$1$^{m}$48$.^{s}$02, $-$36\DEGREE57$'$22\FARCS4)
as CXOU J190148.0-365722 and the positions of the individual detections differed by 
0\FARCS09 for the former and 0\FARCS075 for the latter in RMS.
These sources are separated by only $\sim$0\FARCS8$\pm$0\FARCS12 (1$\sigma$), corresponding 
to a $\sim$140~AU projected separation at $d =170$~pc.

A composite image of 
CXO$_{001007}$, CXO$_{050808}$, CXO$_{050812}$ and CXO$_{050813}$
clearly showed two peaks at the north-east and south-west sides (Figure~\ref{fig:imgcomposite}).
These X-ray peaks have corresponding cm radio counterparts 
within $\sim$0\FARCS1 for each \citep{Choi2008}.
They have offsets from the infrared sources IRS~5a and IRS~5b of $\sim$1\ARCSEC\ 
according to their absolute coordinates in the $H$ and $K$ band maps in Figure 1 of
\citet{Nisini2005}.
The 2MASS survey with an absolute astrometric accuracy of $\sim$100~mas
detected an infrared source very close to the north-east X-ray source
though it did not resolve the two infrared sources probably due to the limited spatial resolution of the CCD (2\ARCSEC).
Moreover,
the two X-ray sources and the IR sources in \citet{Nisini2005} have similar position 
angle and separation, and therefore match their positions well by shifting the IR image
frame (see Figure~\ref{fig:imgcomposite}).
We therefore identified the south-west X-ray source (CXOU J190148.0-365722) as IRS~5a and
the north-east source (CXOU J190148.1-365722) as IRS~5b.

\begin{figure*}
\begin{center}
\plotone{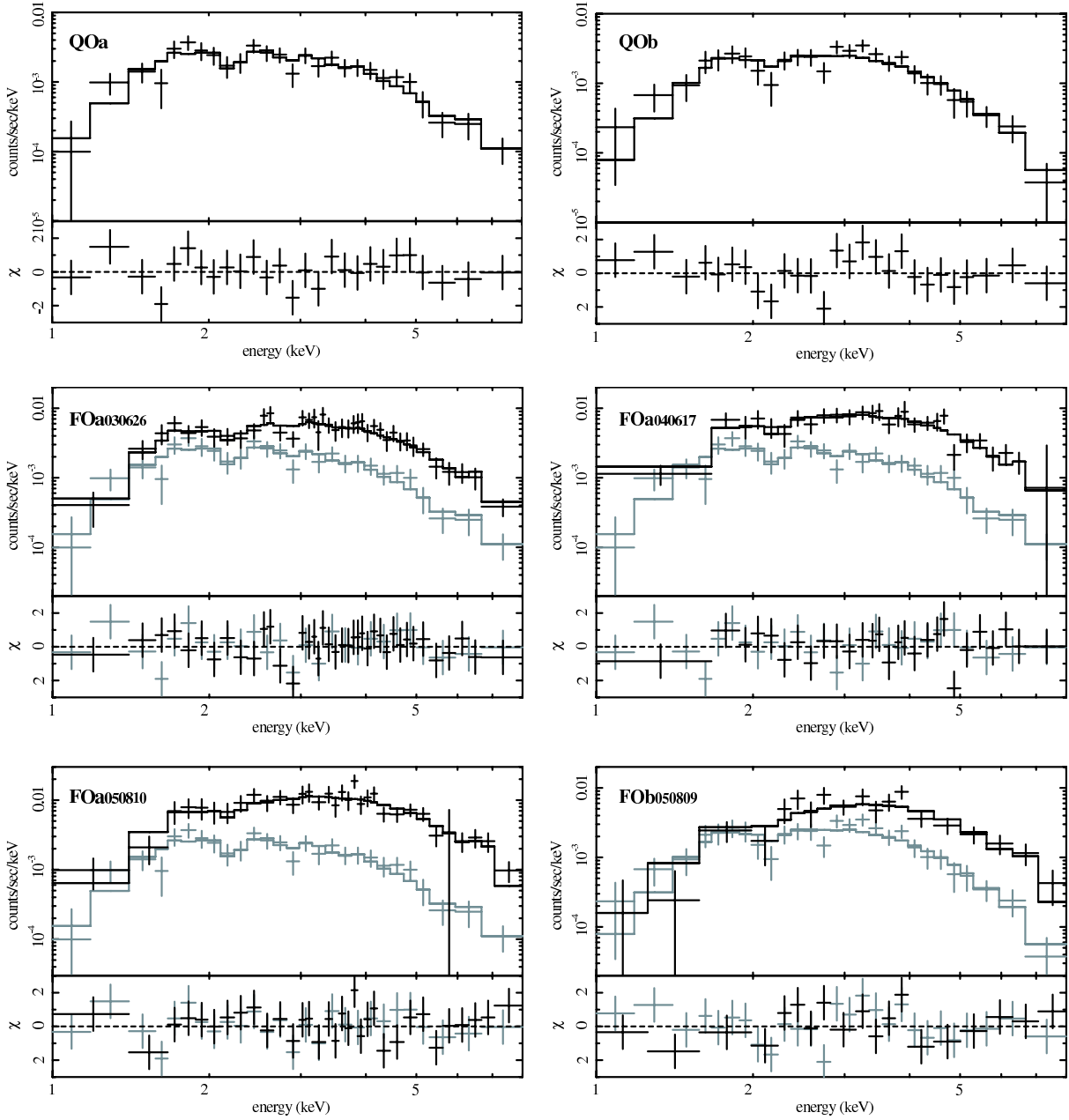}
\caption{Quiescent spectra of IRS~5a ({\it top left}) and IRS~5b ({\it top right}) and flare spectra of IRS~5a
in CXO$_{030626}$ ({\it middle left}), CXO$_{040617}$ ({\it middle right}), CXO$_{050808}$ ({\it bottom left}) and of IRS~5b in CXO$_{050809}$ ({\it bottom right}), overlaid on the quiescent spectra (grey).
\label{fig:spectra}}
\end{center}
\end{figure*}

\begin{deluxetable}{lccccccc}
\tablecolumns{8}
\tablewidth{0pc}
\tabletypesize{\scriptsize}
\tablecaption{Spectral Fits\label{tbl:specres}}
\tablehead{
\colhead{Source}&
\colhead{Spectrum}&
\colhead{Photon}&
\colhead{Model}&
\colhead{\KT}&
\colhead{\NH}&
\colhead{$\chi^{2}$/d.o.f (d.o.f.)}&
\colhead{\LX}\\
&&\colhead{(counts)}&&\colhead{(keV)}&\colhead{(10$^{22}$~\UNITNH)}&&\colhead{(10$^{30}$~\UNITLUMI)}
}
\startdata
IRS5a & QOa & 490.6& 1T & 2.2~(1.7$-$3.2) & 3.7~(3.0$-$4.5)&0.78 (24)&1.5\\
	   & FOa$_{\rm 030626}$ &771.9& 1T~(+QOa)& 5.4~($>$2.8) & 4.7~(3.3$-$6.8)&0.58~(64)&2.5\\
	   & FOa$_{\rm 040617}$ &554.1& 1T~(+QOa)& 5.4~($>$2.8) & 5.1~(3.7$-$6.8)&0.67~(52)&4.4\\	
	   & FOa$_{\rm 050810}$ &597.5& 1T~(+QOa)& 6.6~($>$3.4) & 5.1~(4.1$-$6.5)&0.69~(57)&6.2\\	
IRS5b & QOb & 493.0& 1T & 2.0~(1.5$-$2.5) & 4.5~(3.7$-$5.8)&0.92~(24)&1.9\\
	   & FOb$_{\rm 050809}$ &292.0& 1T~(+QOb)& 4.4~($>$1.4) & 7.8~(5.1$-$14.4)&0.89~(43)&3.3\\
\enddata
\tablecomments{
QO: Quiescent Observation. FO: Flare Observation.
Photon: net photon counts between 1$-$8~keV.
Absorption corrected \LX\ between 0.5$-$10~keV,
assuming the distance of 170~pc.
Values in parentheses denote formal 90\% confidence limits,
and do not include any systematic errors produced by our 2-dimensional fitting procedure.
The elemental abundance was fixed at 0.3 solar value.
}
\end{deluxetable}

\subsection{Timing and Spectral Analysis with 2-Dimensional Image Fits}
\label{subsec:timspec}

Because the X-ray emission from IRS~5a and IRS~5b is strongly blended,
the standard method to generate light curves and spectra --- 
counting photons 
within user defined source and background regions --- cannot be used straightforwardly.
Therefore
we utilized the 2-dimensional image fitting method in which we generated multiple images from each observation in restricted time and energy ranges and fit each 
2-dimensionally by the combined individual model \PSF\ images for IRS~5a and IRS~5b.
We created images
with a size of 80~pixel~$\times$~80~pixel
binned by 0\FARCS123~pixel$^{-1}$.
For the model images, we used the {\tt ptsrc2d} function in the CIAO package,
assuming two point source components at the position of the X-ray sources,
and used the {\tt sherpa} tool to carry out the fitting procedure
(We should note that a part of the X-ray emission may come from different 
and/or extended regions.)
We did not include background in the model images, since the background is negligibly small
($\sim$ 3$\times$10$^{-4}$~cnts / (1 pixel=0\FARCS123)$^{2}$ / 10~ksec).
The {\tt ptsrc2d} function loads an image as a template model \PSF.
Using these model PSFs, we generated photon events at the position of IRS~5
using both {\tt ChaRT} \citep{Carter2003} and {\tt MARX}.
For the light curve analysis, we assumed a \PSF\ to have the typical spectrum
of IRS~5 based on our preliminary analysis using the standard method, 
i.e. \NH\ =4.5$\times$10$^{22}$~\UNITNH, \KT\ =3~keV and 0.3~solar elemental abundance.
In both the timing and spectral analyses,
we generated  \PSF\ images in the same energy band as the observed images
using large numbers of photon events to minimize statistical fluctuations.
In the  {\tt sherpa} fits,
we fixed the centroids of two point source components at the X-ray source positions of 
IRS~5a and IRS~5b and only allowed their normalization parameters to vary.
We used the Cash statistic appropriate for Poisson-limited data and used the Powell method to find the
best-fit values.
We derived upper and lower boundaries of the 1$\sigma$ confidence range 
using the {\tt projection} function.

Figure~\ref{fig:lightcurves} shows individual light curves of IRS~5a and IRS~5b 
in 4~ksec bins in the energy range 1$-$8~keV.
We normalized the individual lightcurves to the 3~keV effective area at the position of IRS~5a,
resulting in small correction factors ($\lesssim$1~\%).
Their flux variations look to be uncorrelated, suggesting no significant interference
with each other in our 2-dimensional fits.
To further check the consistency of the image fitting results,
we  summed the individual light curves of IRS~5a and IRS~5b 
and compared them to a light curve extracted from the entire IRS~5 complex, including photon
counts within a 3\ARCSEC\ radius circle 
centered on between IRS5a and IRS5b, which 
include $\sim$90$-$93\% of photons from both IRS~5a and IRS~5b.
We found that the standard method obtained 93$-$98\% of the photon counts 
we derived by combining the individual IRS~5a and IRS~5b photons.
We had similar results for the split energy bands between 1$-$3.5~keV and 3.5$-$8~keV,
which we used for a hardness ratio analysis described in the next paragraph.
This suggests that our image fitting method produces a result that is consistent with the standard method
within $\lesssim$10\% discrepancy.

In CXO$_{030626}$, CXO$_{040617}$, and CXO$_{050810}$,
IRS~5a showed significant time variations, reminiscent of stellar X-ray flares.
The fast rise ($\sim$4~ksec in CXO$_{040617}$) and 
slow decay (a half decay time of 15$-$30~ksec) time scales based on visual inspection
were similar to those of relatively strong X-ray flares from YSOs.
The flux level did not change remarkably during the other observations ($\lesssim$50\%).
IRS~5b was quite stable in almost all the available observations.
The non-detection of IRS~5b with {\tt wavdetect} in 4 observations would be therefore
caused by increased contamination from IRS~5a by flares, or from limited photon statistics.
IRS~5b showed a small flux enhancement by a factor of 2 in CXO$_{050809}$,
and an apparent decrease at the middle of CXO$_{040617}$, but this decline 
could be an artifact produced by strong contamination from the IRS~5a flare 
in this observation along with a possible offset of
the south-west emission peak from IRS~5a.
We split the energy band into two at 3.5~keV, which is near the median of 
the photon distribution against energy in a spectrum, 
and measured the hardness ratio ($HR$) of each observation,
which we have defined as $HR = (H-S)/(H+S)$,
where the $H$ and $S$ are count rates of the 3.5$-$8~keV and 1$-$3.5~keV bands, respectively.
The average $HR$s of both IRS~5a and IRS~5b
in CXO$_{001007}$, CXO$_{050808}$, CXO$_{050812}$ and CXO$_{050813}$,
when neither showed significant variations
(and which we refer to as the ``quiescent observations''),
were $HR\sim-$0.32$\pm$0.15, 
$-$0.29$\pm$0.18 (1$\sigma$), respectively.
The $HR$s were $\sim-$0.02$\pm$0.09 (1$\sigma$) during flares from IRS~5a 
(CXO$_{030626}$, CXO$_{040617}$, CXO$_{050810}$)
and $\sim$0.04$\pm$0.08 (1$\sigma$) during a small flux enhancement of IRS~5b
(CXO$_{050809}$).
The $HR$ was slightly higher in the flare than during the quiescent observations.

We generated individual spectra of IRS~5a and IRS~5b in key phases 
and analyzed them with {\tt xspec} version~11.
One technical issue was that {\tt xspec} requires integer photon counts,
while the {\tt ptsrc2d} function outputs the normalization values (i.e. photon counts) in floating point.
Thus, to avoid any round off error,
we multiplied photon counts and exposure time by 1000 when creating spectral files.
The {\tt xspec} format also assumes a single error value for each spectral bin,
while the {\tt projection} function outputs upper and lower error ranges.
We therefore used the larger value output from {\tt projection} as the {\tt xspec} error for each spectral bin.  Since 
we set most of the spectral bins to have $\gtrsim$15~photons
the upper and lower error ranges typically have less than $\sim$10\% differences.
We generated response matrices and auxiliary files corresponding to the spectral files
using {\tt mkacisrmf} and {\tt mkarf} in the CIAO package.

Since no quiescent observation had
enough photon statistics to generate a spectrum with sufficient signal-to-noise,
we combined images of these observations in each spectral bin.
We combined their \PSF\ images and spectral responses as well for our 2-dimensional 
image fits and spectral model fits.
The result is shown in Figure~\ref{fig:spectra}.
The quiescent spectra of IRS~5a and IRS~5b were almost identical with
significant emission up to $\sim$8~keV and a cut-off due to absorption at $\sim$1.5~keV.
We modeled both spectra using an absorbed optically thin, 1-temperature (1T) plasma model 
(WABS, \citealt{Morrison1983}; APEC\footnote{http://cxc.harvard.edu/atomdb/}, 
Table~\ref{tbl:specres}).
The best-fit models had \NH\ $\sim$4$\times$10$^{22}$~\UNITNH\ and \KT\ $\sim$2~keV
for both sources.
The absorption corrected log \LX\ was 30.2$-$3 (0.5$-$10~keV).

We compared the IRS~5a spectra during flaring states
(CXO$_{030626}$, CXO$_{040617}$ and CXO$_{050810}$)
to its quiescent spectrum (Figure~\ref{fig:spectra}).
All flare spectra showed stronger emission in the 1$-$8~keV energy band
than the quiescent spectrum.
We fit the flare spectra by an absorbed 1T plasma model plus the best-fit model of
the quiescent IRS~5a spectrum, assuming that a new plasma component appeared
during the flares.
The best fit to the flare spectra seems to have higher \KT\ 
($\sim$5$-$7~keV) than the quiescent spectrum but  similar \NH\ ($\sim$5$\times$10$^{22}$~\UNITNH).
However,
IRS~5b showed similar increase in hard band flux in CXO$_{050809}$ 
compared to the quiescent phase, while the flux below $\sim$2~keV was apparently 
unchanged (Figure~\ref{fig:spectra}).
The best fit, assuming an absorbed 1T model plus the best-fit model of
the quiescent IRS~5b spectrum showed significantly larger 
\NH~$\sim$7.8$^{+4.2}_{-1.6}\times$10$^{22}$~\UNITNH\ (1$\sigma$),
which does not overlap with \NH\ in the quiescent spectrum (4.5$^{+0.4}_{-0.5}\times$10$^{22}$~\UNITNH, 1$\sigma$).
The best-fit model also showed slightly higher \KT\ $\sim$4~keV than the quiescent spectrum.
None of these spectra had enough photon statistics to show the Fe line profile,
which appeared in some flare spectra of the entire IRS~5a complex.

The values of \NH\ and \KT\ for IRS~5a and IRS~5b are similar 
in all the observations and are consistent with the values in \citet{Forbrich2006},
which were derived from the combined spectra of IRS~5a and IRS~5b.
The \KT\ and \LX\ values are typical of those from TTSs and Class~I protostars,
while the derived \NH\ is typical of Class I protostars \citep[e.g.,][]{Imanishi2001}.

\begin{deluxetable}{ccccccc}
\tablecolumns{7}
\tablewidth{0pc}
\tabletypesize{\scriptsize}
\tablecaption{Young Binary System with $<$200~AU Projected Separation 
Resolved in X-rays \label{tbl:xraybin}}
\tablehead{
\colhead{Object}&\colhead{System}&\multicolumn{2}{c}{Separation}&\colhead{Distance}&\colhead{Age}&\colhead{Reference}\\ \cline{3-4}
&&\colhead{sky}&\colhead{projection}\\
&&\colhead{(arcsec)}&\colhead{(AU)}&\colhead{(pc)}&\colhead{(Myr)}
}
\startdata
TWA 5 & M3Ve (TTS) + M8.5$-$9 (BD)&2&110&55&12&\citet{Tsuboi2003}\\
HD~100453&A9Ve + M4$-$M4.5 (TTS) &1.06&114&120&20&Collins et al. in prep.\\
HD~98800 &K5Ve (TTS) + ?&0.8&48&38&5$-$10&\citet{Kastner2004b}\\
R CrA IRS~5&K5-K7V (Class~I) + ? &0.8&140&170&0.3$-$0.5 or 1.3\tablenotemark{a}&this result\\
\enddata
\tablecomments{TTS: T-Tauri Star, BD: Brown Dwarf.
}
\tablenotetext{a}{Values of IRS~5a from \citet{Nisini2005}.}
\end{deluxetable}

\section{Discussion}
\label{sec:discussion}

We still have a discrepancy between the positions of the near-infrared sources in \citet{Nisini2005} 
and the sources detected by 2MASS, VLA and \CHANDRA.
IRS~5a was a factor of two brighter than IRS~5b in \citet{Nisini2005},
while the 2MASS source, whose position should be biased toward the brighter source (IRS~5a),
was instead closer to IRS~5b (see \S\ref{subsec:anaimage}).
The 2MASS source is unlikely to be mis-positioned due to surrounding contamination 
since both IRS~5a and IRS~5b are sufficiently bright in the images shown in Figure~1 of \citet{Nisini2005}.
IRS~5a or IRS~5b might have experienced an increase in its near-infrared flux
during the observations in \citet{Nisini2005} or during the 2MASS observations.
We should note that
IRS~5 only showed variation of less than $\pm$0.03mag during monitoring observations 
in 2005 \citep[see Figure 6 of][]{Forbrich2007}.

It is possible that the X-ray sources we have identified with either IRS~5a or IRS~5b could be a background red giant star or an active galactic nucleus, though this is unlikely.
However,
X-ray emission from red giant stars are not usually as bright 
as, nor as hot as, emission from IRS~5a or IRS~5b \citep{Pizzolato2000}.
Though some AGNs are observed with fluxes similar to those from IRS~5a or IRS~5b, 
the X-ray spectra are generally harder 
\citep[spectral photon index $\Gamma \lesssim$2, ex.][]{Ueda1998}
than those observed for either IRS~5a or IRS~5b ($\Gamma \sim$3.5 assuming an 
absorbed power-law model).
On the other hand,
\citet[][]{Pontoppidan2003} showed CO ice absorption lines of IRS~5a and IRS~5b 
have similar systemic velocities to the other YSOs in the R~CrA cloud ($\sim-$20~\UNITVEL)
though this feature can be produced by an absorber physically unrelated to the infrared sources.
The X-ray spectra are typical X-ray emission from Class~I protostars.
These results suggest that IRS~5a and IRS~5b form a protostar binary system within the R CrA cloud.

Table~\ref{tbl:xraybin} shows binary systems with projected separations
less than 200AU that have been spatially resolved in X-rays.
According to this table, IRS~5 is the first X-ray binary system believed to be comprised of Class~I 
protostars\footnote{Although the infrared spectral slope, 
which defines the YSO classification,
has not been measured for IRS~5b, 
we think this star is a Class~I protostar based on its high
absorption in X-rays.
}
and is a factor of $\gtrsim$10 younger than the other systems.
Binary systems with linear separations of less than 140~AU have very little disk mass,
perhaps because of the gravitational influence of the binary companion \citep{Beckwith1990}.
The IRS~5 binary separation is near this boundary and the separation may be smaller 
if the distance is $d\sim$130~pc.
The separation is so large that the binary stars would not have direct interaction
by magnetic field, for example 
(see \citealt{Uchida1985} as an example of the system with magnetic interaction),
but they may have an influence as discussed by \citet{Beckwith1990}.
IRS~5 may be an important example of the influence on X-ray activity of binary
interaction in a very early phase.

The column densities of IRS~5a and IRS~5b derived from their X-ray spectra
($\sim$4$\times$10$^{22}$~\UNITNH)
were typical of Class~I protostars \citep[c.f.][]{Imanishi2001}.
These values of \NH\ were a factor of 2 smaller than the \NH\ value estimated from \AV\ for IRS~5a ($\sim$45~mag), 
assuming an empirical \NH$-$\AV\ relation appropriate to the $\rho$ Oph cloud \citep{Imanishi2001}.
This result was consistent with X-ray absorption measurements 
of IRS1, 2 and 5 \citep{Forbrich2006} and 
qualitatively consistent with a small \NH/A$_{\rm J}$ ratio of the R CrA cloud \citep{Vuong2003}.
The moderate \NH\ may not suggest the edge-on geometry discussed by \citet{Nisini2005}
since X-rays from the stellar core would suffer stronger extinction in the edge-on disk
than infrared emission from the stellar core and inner disk.

The \KT\ and \LX\ values of the quiescent spectra of both IRS~5a and IRS~5b 
are typical of Class~I protostars and TTSs \citep{Imanishi2001}.
Assuming the mass of IRS~5a between 0.4$-$0.9~\UNITSOLARMASS\ \citep{Nisini2005},
the \LX\ $\sim$10$^{30.2}$ \UNITLUMI\ corresponds to a  $\sim$1~M-year-old pre-main sequence star
based on the \LX-age relation of young stars in the Orion nebula though the relation has a large scatter \citep{Preibisch2005}.
The \KT s during the flares of IRS~5a are typical of YSOs
\citep[e.g.,][]{Imanishi2001,Wolk2005}.

IRS~5a exhibited 3 prominent flares during the 8 \CHANDRA\ observations.
Since the observations ended at the middle of these flares,
we only know lower limits to the flare durations ($\gtrsim$37.6, 19.9 and 15.0 ksec)
and total flare energies ($\gtrsim$9.2, 8.7 and 9.3 $\times$10$^{34}$~ergs),
respectively.
These flares were similar in brightness to that of the brightest flares  
from classical TTSs and weak-line TTSs
observed in the XEST project \citep[Fig. 9 of][]{Stelzer2007}. 
Curiously, we did not see any flare-like event from IRS~5a fainter than those flares 
though such flares would have been detectable.
IRS~5a had 3 flares during the 152~ksec exposure, that is, one flare per $\sim$50~ksec.
This is more than an order of magnitude more frequent than the average of the XEST 
 \citep[1 flare in 770~ksec for flares with the flare energies above 10$^{35}$~ergs,][]{Stelzer2007}
and COUP \citep[1 flare in 650~ksec,][]{Wolk2005},
though we did not consider possible detections of flares that start outside the observing window.
IRS~5 showed remarkable variation in the other \XMM\ observations too \citep{Forbrich2006},
for which either of IRS~5a or IRS~5b would be responsible.
IRS~5a seems to be an actively flaring source,
though the accretion rate was measured to be low \citep{Nisini2005}.

IRS~5b did not show apparent variation during the \CHANDRA\ observations
except for a flux enhancement in CXO$_{050809}$.
The column density of IRS~5b during the flux enhancement
increased by a factor of $\sim$2 from the quiescent phase.
Such variable X-ray emission in CXO$_{050809}$ might have been produced from a plasma
on or around the stellar surface that appeared behind the circumstellar disk,
or perhaps IRS~5b may have another deeply embedded companion.

Except for the flare activity,
X-ray properties of IRS~5a and IRS~5b are strikingly similar.
This may strengthen a general assumption that stars in a similar mass range experience 
similar evolution of their quiescent X-ray spectra.
This result also raises a caution in  measurements of the \LX\ function of YSOs ---
\LX\ can be significantly overestimated if a YSO has an unresolved companion.

\section{Conclusion}

We spatially resolved X-ray emission from each component of the infrared double system 
IRS~5 in the R CrA molecular cloud with a 0\FARCS8 projected separation in 8 \CHANDRA\ observations.
Their X-ray brightnesses and spectra suggest that
they are a Class~I protostar binary system.
Separations of binary YSOs are typically less than $\sim$10\ARCSEC\ in nearby star 
forming regions.
\CHANDRA\ enabled study of X-ray properties of such young binary systems
for the first time.

We derived light curves and spectra of these objects using a 2-dimensional image
fitting method.
During the 8 \CHANDRA\ observations,
the southern source IRS~5a showed three X-ray flares lasting more than
15~ksec and with flare energies of $\gtrsim$10$^{35}$ ergs.
These flares were relatively powerful compared to flares from classical TTSs and weak-line TTSs.
The flare frequency (1 flare per $\sim$50~ksec) was higher than flares from 
PMSs in the Taurus and Orion cloud.
IRS~5a might be in an active phase, or may have a different mechanism which drives its X-ray flares.
IRS~5b, on the contrary, did not show strong variation through the observations,
except for a flux increase by a factor of two in 2007 Aug. 9 accompanied by
an apparent \NH\ increase.

Though IRS~5a and IRS~5b had clear differences in their X-ray variability,
their quiescent X-ray spectra were almost identical.
This may suggest that both stars have gone through a similar evolutionary phase
in X-ray activity.
Detailed infrared spectroscopy of IRS~5b is crucial to 
determine its stellar parameters and to understand nature of the IRS~5 system.
Measurement of proper and radial motions of both IRS~5a and IRS~5b
would be also important to understand whether they are a gravitationally bound system.

\acknowledgments

This work is performed while K.H. was supported by the NASA Astrobiology Program 
under CAN 03-OSS-02.
This publication makes use of data products from the Two Micron All Sky Survey, which is a joint project of the University of Massachusetts and the Infrared Processing and Analysis Center/California Institute of Technology, funded by the National Aeronautics and Space Administration and the National Science Foundation.
This research has made use of data obtained from the High Energy Astrophysics Science Archive Research Center (HEASARC), provided by NASA's Goddard Space Flight Center.

Facilities: \facility{CXO(ACIS-I)}

\bibliographystyle{apj}
\bibliography{inst,sci_AI,sci_JZ,scibook}

\end{document}